\documentclass[conference]{IEEEtran}
\IEEEoverridecommandlockouts
\usepackage{cite}
\usepackage{amsmath,amssymb,amsfonts}
\usepackage{algorithmic}
\usepackage{graphicx}
\usepackage{textcomp}
\usepackage{xcolor}
\def\BibTeX{{\rm B\kern-.05em{\sc i\kern-.025em b}\kern-.08em
    T\kern-.1667em\lower.7ex\hbox{E}\kern-.125emX}}
\begin{document}

\title{TurboLoRa: enhancing LoRaWAN data rate via device synchronization}

\author{\IEEEauthorblockN{Moez Altayeb}
\IEEEauthorblockA{\textit{Marconi Lab} \\
\textit{ICTP}\\
Trieste, Italy \\
mohedahmed@hotmail.com}
\and
\IEEEauthorblockN{Marco Zennaro}
\IEEEauthorblockA{\textit{Marconi Lab} \\
\textit{ICTP}\\
Trieste, Italy \\
mzennaro@ictp.it}
\and
\IEEEauthorblockN{Ermanno Pietrosemoli}
\IEEEauthorblockA{\textit{Marconi Lab} \\
\textit{ICTP}\\
Trieste, Italy \\
ermanno@ictp.it}
\and
\IEEEauthorblockN{Pietro Manzoni}
\IEEEauthorblockA{\textit{Computer Networks Research Group} \\
\textit{Universitat Polit\`ecnica de Val\`encia}\\
Val\`encia, Spain \\
pmanzoni@disca.upv.es}
}

\maketitle

\begin{abstract}
Over the last few years we have witnessed an exponential growth in the adoption of LoRaWAN as LPWAN technology for IoT. While LoRaWAN offers many advantages, one of its limitations is the paltry data rate. Most IoT applications don't require a high throughput but there are some that would benefit from a higher data rate. 

In this paper, we present TurboLoRa, a system that combines the strengths of LoRaWAN while providing a higher data rate by synchronizing the transmission of multiple LoRaWAN devices. Our proposal allows to combine cheap devices making it a frugal solution to this kind of problems. We present some  preliminary results obtained using a real prototype of TurboLoRa.
\end{abstract}


\section{Introduction}
In the last few years we have observed a very fast growth of LoRaWAN deployments. Community networks such as TheThingsNetwork\footnote{https://www.thethingsnetwork.org/} (TTN) have fostered the adoption of LoRaWAN worldwide. One of the strengths of LoRaWAN is that it allows for very long links while providing a strong security mechanism and requiring little power. The biggest limitation of this technology is the low data rate. 

The amount of data that IoT devices collect and transmit will increase as the technology continues to develop, which will in turn contribute to the need for increased bandwidth. Consumers expect that bandwidth will always be available at the fastest speeds possible, even as IoT demand grows~\cite{8241454}.

Increasing the available data rate would allow applications with demanding payloads like simultaneous output from several sensors, images or audio  to be deployed \cite{8909245,8958675}. The use of images is expected to greatly increase, stimulating broader use of unmanned (e.g., drone and robotic) inspection and monitoring applications. For example, drones will autonomously survey crops in the field using visual analytics running near the edge to assess factors including growth rates and pest losses. This information could allow farmers to optimize harvest time and more efficiently target particular areas for pesticide, fertilization, irrigation or other interventions. Many other IoT based applications can take advantage of increased throughput by developing edge solutions like remote-controlled robotics, near real-time video analysis and other kinds of factory-floor automation.

LoRaWAN is a standard based on the proprietary LoRa modulation that adds the required layers for a complete end to end secure communication. 
It uses a pure ALOHA media access mechanism. Collision avoidance  relies on a low channel occupancy, which suits the need of many IoT applications, but fails short in the support of those demanding higher throughput and very low latency, because of the duty cycle or maximum dwell time restrictions. The pseudo orthogonality of the different spreading factors allows for simultaneous reception at the same frequency, but in practice this only works when the difference in power among the competing signals is bounded \cite{8267219}, so in this work we only use the shortest spreading factor since we want to maximize the data transfer rate.   

We present TurboLoRa, a system that combines the strengths of LoRaWAN while providing a higher data rate by synchronizing the transmission of multiple LoRaWAN devices. Our proposal allows to combine cheap devices making it a frugal solution to this kind of problems~\cite{fru8048444}. Some  preliminary results obtained using a real prototype are offered, too.

The paper is organized as follows. 
In Section \ref{sec:lorawan} we briefly provide some details of the LoRaWAN technology.
Section \ref{sec:turbolora} describes the details of the proposal, 
and Section \ref{sec:evaluation} present the preliminary  results obtained with a prototype. Finally, Section \ref{sec:conclusions} present some final comments and ideas to be tackled in future work.

\section{LoRaWAN}
\label{sec:lorawan}
LoRaWAN \cite{8030482} uses unlicensed frequencies (centered at 868 MHz  or 912 MHz), so a mechanism to share the channel among different users must be deployed. In ITU Region 1, comprising Europe and Africa, a device must not transmit more than 1  \%\ of the time, while in ITU Region 2, comprising the Americas, there is a limit on the maximum dwell time of 400 ms, after which the transmission must cease to accommodate  other possible users. These restrictions apply to a specific frequency, so by hopping to a different one new transmission opportunities arise. 
LoRa modulation  uses chirps of varying frequencies, occupying the whole available channel bandwidth. The amount of spectrum spreading  is controlled by the so called spreading factor, SF, which in Region 1 can have values between 7 and 12. Shorter SFs produce shorter frames, occupying less air time and allowing for greater throughput, but require higher signal to noise ratios for reception. Contrariwise, longer SFs can be detected even when the signal is much weaker than the noise, capable not only of reaching longer distances but also to overcome the attenuation of walls and floors, as well as that introduced by vegetation encountered in the transmission path. Signals with longer SF will occupy the channel for a considerable longer time. This increases the probability of collisions, which will raise the packet error rate, the key performance parameter in many applications.
So the SF allows tailoring  each frame to the particular conditions of a channel.  As an example, SF 12 can decode signals that are 100 times weaker than the noise or interference in the channel.

\section{TurboLoRa}
\label{sec:turbolora}
In this Section we describe TurboLora, a solution we developed to address the data rate limitation while maintaining  the functionalities of LoRaWAN, namely: authentication, end to end encryption, error detection and correction. The architecture is the same star-of-stars topology as that of LoRaWAN \cite{8502812}.

TurboLoRa is based on the simultaneous transmission  of several LoRa devices  commanded by a unique synchronization signal, so that the throughput is multiplied by the number of devices. The end node splits the data among several LoRa devices that transmit them in parallel. As in standard LoRaWAN, data are received by one or more Gateways that forward them to a Network Server; from here data can be accessed by the application using any  Internet connection.
It is also possible to incorporate the Gateway, Network, and Application servers in a single box, providing a completely autonomous solution.
As the data are sent in parallel, different channels have to be used to avoid collisions.  This fact does not introduce any problem since a full-fledged  LoRa Gateway must be able to listen simultaneously to eight  frequencies to receive any possible transmission.

The complete payload, which could be an image or an  audio file,  is sent by the source node to the LoRa devices divided in equal pieces so that each device in the system sends the same amount of data. The data is sent to each LoRa device via a serial connection. Once the data is transferred, a synchronization signal commands the devices to send simultaneously their part of the payload using different frequencies.

\begin{figure}[ht]
  \centering
  \includegraphics[width=\linewidth]{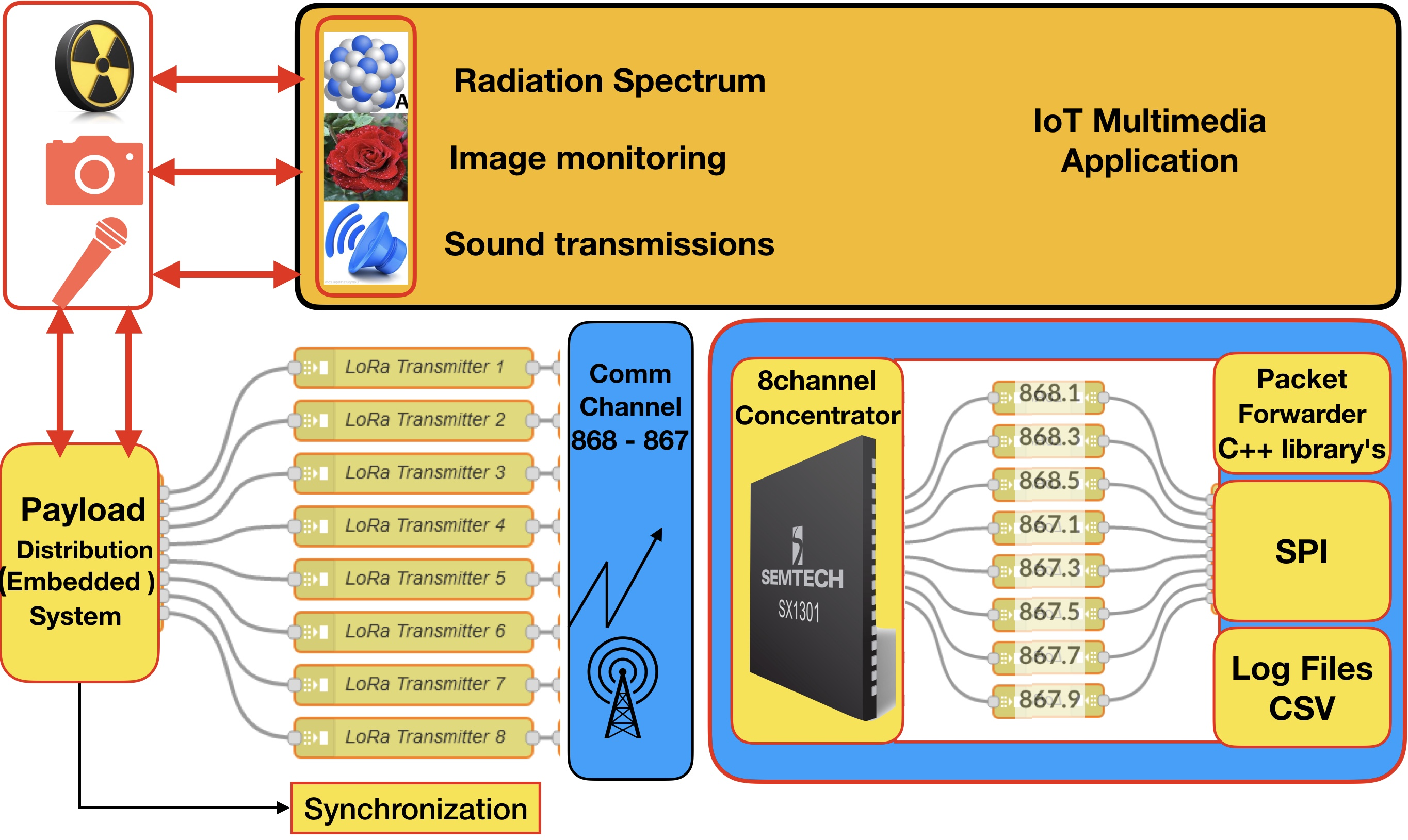}
  \caption{Leveraging eight LoPy devices in TurboLora to reduce transmission time.}
  \label{fig:ducting1}
\end{figure}

\begin{figure}[ht]
  \centering
  \includegraphics[width=\linewidth]{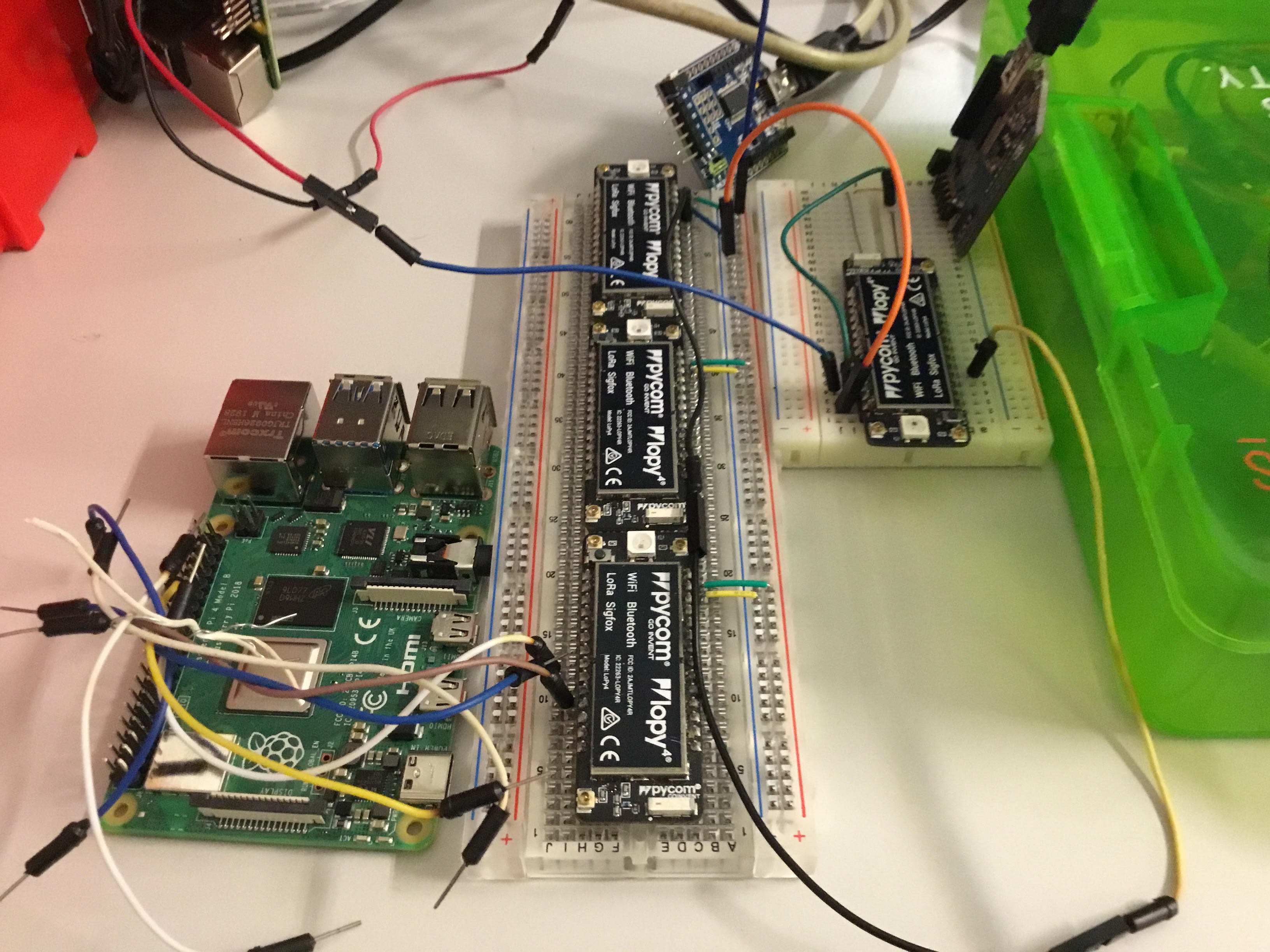}
  \caption{Prototype end node using four LoPy and one Raspberry Pi.}
  \label{fig:ducting2}
\end{figure}

\begin{figure}[ht]
  \centering
  \includegraphics[width=\linewidth]{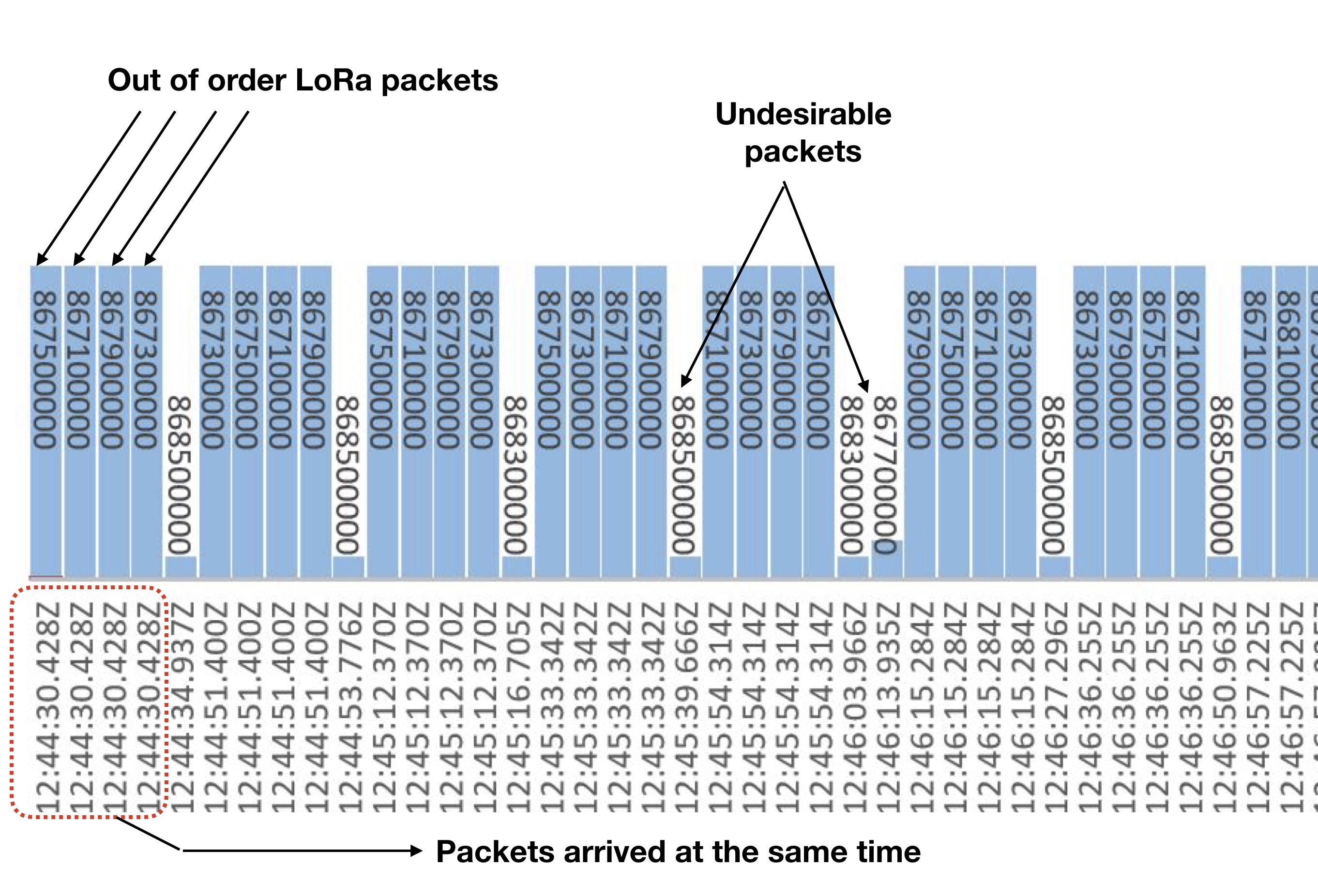}
  \caption{Received packets frequencies versus time, showing both TurboLora packets out of order and normal LoRaWAN packets from other applications.}
  \label{fig:ducting3}
\end{figure}

Figure \ref{fig:ducting1} shows a generic architecture of a TurboLoRa node with eight devices, and 
Figure \ref{fig:ducting2} shows our prototype end node. We used four LoRa devices as a proof of concept. These devices are  LoPys  by Pycom\footnote{https://pycom.io/}, which are a system-on-a-chip product based on the ESP32 microcontroller which offers integrated LoRa, Sigfox, WiFi, Bluetooth and NB-IoT connectivity options. It has several inputs that can be used to send signals to the microcontroller. The two UARTs can be used for serial communications and there are up to 24 General-Purpose Input Output (GIO) pins that can be used for synchronization. The LoPy has 4 MB of RAM and can be programmed in MicroPython.
The four LoPy devices are connected to as Raspberry Pi which sends the data via serial connection and the synchronization signal via the GPIO.
As an example, Figure~\ref{fig:ducting3}  illustrates the  packets reception on 4 different frequencies at exactly the same time. 

\subsection{TurboGateway}

To leverage the security features of LoRaWAN and to use the same gateway for other ``normal'' LoRaWAN end nodes, we developed the code to reassemble the packets coming from the end node. Data chunks arrive at the gateway out of order, as shown in Figure~\ref{fig:ducting3}. We integrate the Packet Forwarder utility code in the IMST Lite Gateway based on the iC880A concentrator with special Python functions. The code extracts the data from the concentrator, filters packets from our nodes out of those coming from other nodes,  and then decodes and reorders the packets. To access the results we implemented our own Network and Application Servers, running  in the same platform that implements the Gateway, but they can also be  visualized through TheThingsNetwork (TTN) \cite{9045213} by means of integration with Ubidots\footnote{https://www.ubidots.com/}.

The maximum LoRa payload for a single transmitter is 255 bytes at spreading factor 7. In this work the payload is set at 226 bytes, sent over 125 kHz of bandwidth with forwarding correction code rate 4/5 (5 bits sent to convey 4 bits of information). This results in a symbol duration of 1.02 ms. Accounting for  the LoRaWAN overhead, we are transmitting 338 symbols per frame plus  symbols for preamble and synchronization, so the total duration is  358.7 ms per frame. To comply with the duty cycle limitation of  1 \% in Region 1, we have to wait at least 35.64 s after each frame transmission. Therefore the LoPy was set to sleep for 36 seconds before the next transmission. 



\section{Evaluation results}
\label{sec:evaluation}
To evaluate the performance of our solution we developed a demo application that sends a simple image via TurboLoRa. The image could be that of a leaf of a plant  to be analyzed using ML (Machine Learning) techniques to extract  features like degree of ripeness, presence of  disease or parasite, health, state of ripeness or irrigation need. The Raspberry Pi in the end node has enough processing power to perform simple  ML processing, while  the LoRa high link power budget can allow for long range and mitigate the attenuation effects of vegetation or other obstacles in the trajectory. 

The Raspberry Pi was fitted with the Pi Camera Module to capture the image using the Open CV software (Open Source Computer Vision Library). After capturing the image, it was resized  to 225x225 pixels, each  with 8  levels of gray. The next step was reshaping the matrix into a vector to be able to divide the data in four parts so that the four LoPys could send them  in parallel. 

\begin{figure}[ht]
  \centering
  \includegraphics[width=\linewidth]{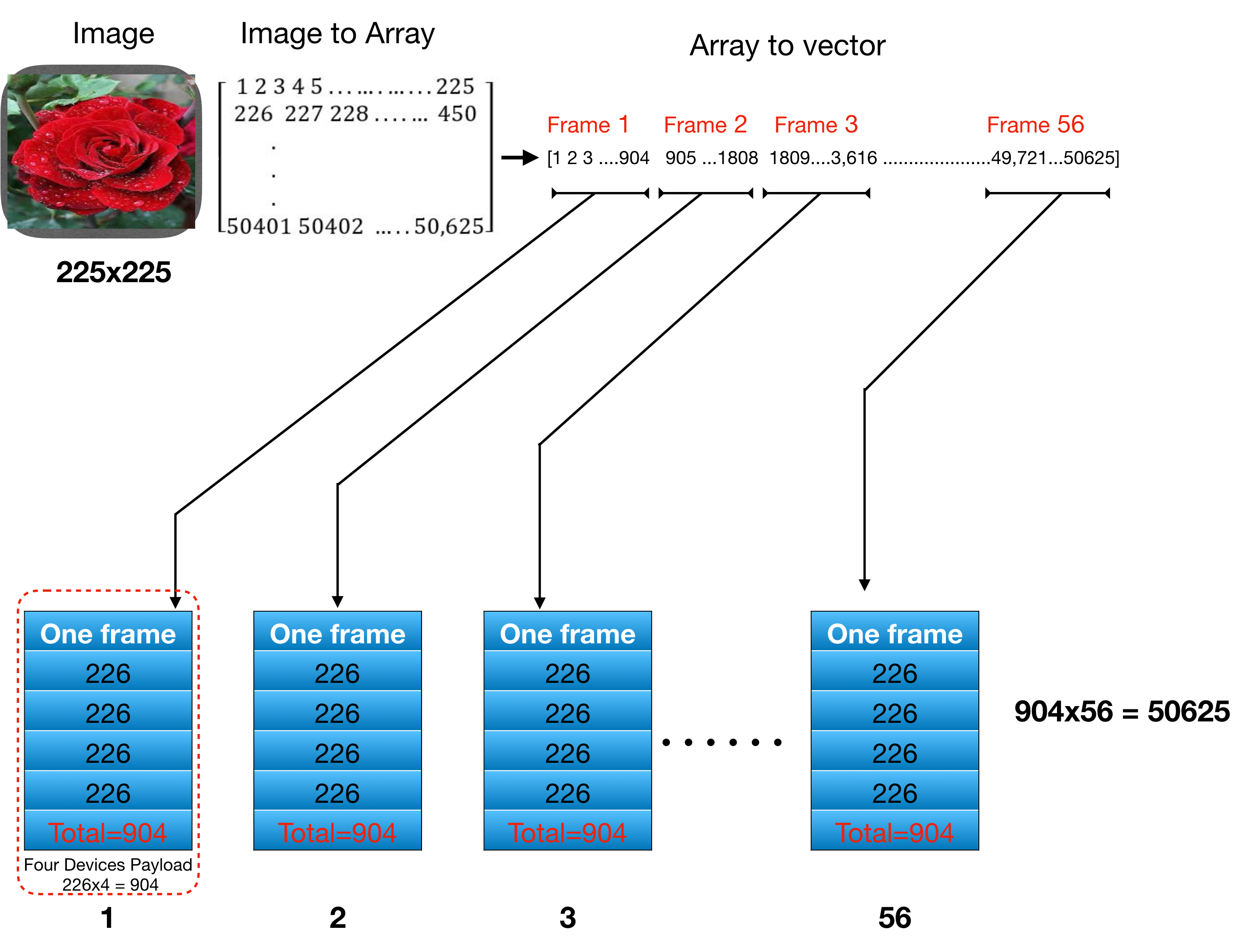}
  \caption{Reshaping the matrix as a vector to divide the data in four parts.}
  \label{fig:ducting5}
\end{figure}

Figure \ref{fig:ducting5} shows the reshaping of the matrix into a vector. The first column of each row has an index, ranging from 1 to 225. This index will be used to reconstruct the matrix once it has been transmitted by the four devices. The numbers of elements in the matrix go from [1] to [50625].
 
The embedded system at this stage converts the vector data into bytes values in order to transfer it to the LoRa controller device. A special serial communications setup is used to match the LoRa controller buffer's serial port and memory capacity. The maximum LoRa payload for a single transmitter is 255 bytes at spreading factor 7. In this work the payload is set to 226 bytes, so the four devices transmit simultaneously  a total of 904 bites. The first LoPy transmits bytes [1:225], the second  [226:450] and so on, until  the fourth which sends bytes [678:904]. After one batch transmission, which occupies the channel for  358.7 ms,  the LoPys are put in deep sleep for 36 seconds, yielding a duty cycle  of 0.99\%, thus adhering to the Europe’s  ETSI EN 300.220-2 V3.2.1 regulations.
This process is repeated 56 times to send the complete image, taking a total time of 2036.16 seconds, that is 34 minutes, so up 40 images per day can be transferred. 
Using one LoRa device, the total time would have been four times longer.

 We used SF7 in the four devices. While this does not allow the highest sensitivity, it shortens the time on air.
 Although we  tested  the prototype with a low resolution image as proof of concept, the extension to a higher resolution image is straightforward applying the same technique of vectorizing the matrix in the transmitter and performing the reverse operation in the receiver.
 Of course, other type of files, like for instance digitized audio, can also take advantage of the TurboLora transmission.



\begin{figure}[ht]
  \centering
  \includegraphics[width=\linewidth]{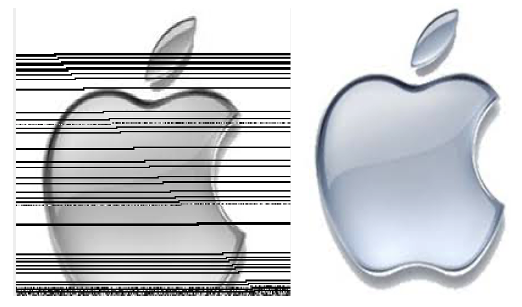}
  \caption{An example of corrupted frames. The image on the right is the original, the one on the left shows corrupted frames as black lines.}
  \label{fig:ducting7}
\end{figure}

After the transmitters send their data, they are received by the gateway possibly unordered. The first operation is to decode the packets and to order them according to the packet number. Once this is done, the payload can be extracted. The data is then converted from bytes to unsigned integers numbers, in order to reconstruct the image.  The vector data is reshaped as a matrix with the dimension of the original transmitted  (225x225). 
Figure \ref{fig:ducting7} shows the original image and the reconstructed  one at the receiver. The black lines are due to  corrupted data detected by the software, which writes  a value of zero whenever it encounters a transmission error, as a result of collision or poor signal level. 


Since in our lab we have many LoRa  nodes transmitting at full power, the prototype  experienced many damaged frames. The software in the gateway waits until the transmission is over and checks the Cyclic Redundancy Check (CRC) flag of each frame. It saves all the bad CRC package numbers in an array and sends the array back to the transmitters. The related transmitter sends the data again. The algorithm is repeated until all bad frames have been recovered.

\begin{figure}[ht]
  \centering
  \includegraphics[width=\linewidth]{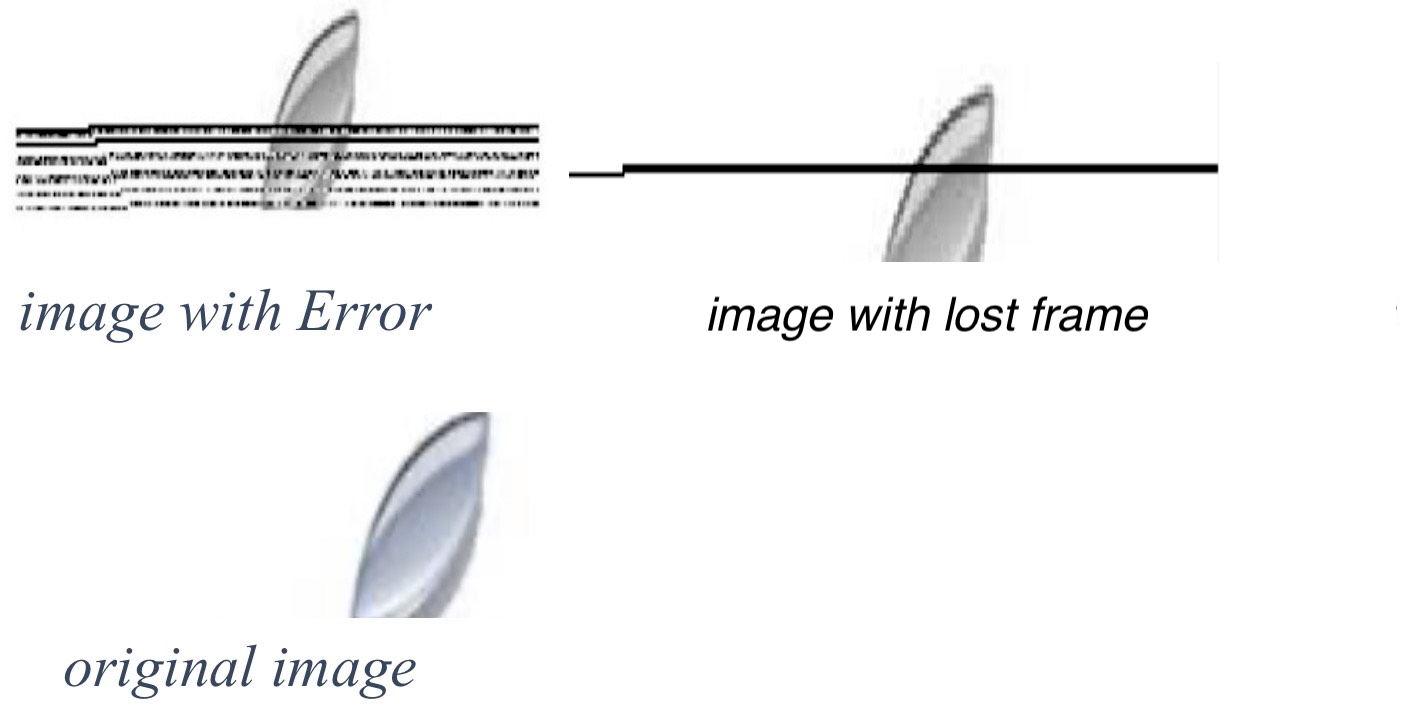}
  \caption{Frames  lost  and frames with errors are shown as black lines in the reconstructed image.}
  \label{fig:ducting86}
\end{figure}

In addition to packets with a damaged frame, which are detected using the CRC, there are also instances of lost frames. A lost frame is acceptable in certain LoRaWAN transmissions where a single measurement is not essential when one is  looking at overall trends (for example in temperature measurements). In the case of image transmission, as in many others in which corrupted data have to be pinpointed, it is important to identify the lost frames in the transmission. The decoding function that we implemented substitutes the lost frames with one filled with 0 values that show as a black lines in the reconstructed image. Figure \ref{fig:ducting86} shows an example of reconstructed images with errors and with lost frames. The original image is shown for comparison.

\section{Conclusions and Future work}
\label{sec:conclusions}

In this paper we presented TurboLoRa, a system that combines the strengths of LoRaWAN while providing a higher data rate by synchronizing the transmission of multiple LoRaWAN devices. Our proposal allows to combine cheap devices making it a frugal solution to this kind of problem. We present some  preliminary results obtained using a real prototype of TurboLoRa. Although our prototype employed only 4 transmitters, the extension to 8 transmitters commanded by a single synchronization signal is straightforward, and would lead to an 8 times improvement of the throughput, allowing for a wealth of new applications that cannot be met with the current LoRaWAN solutions. Applications in agriculture, industry and health are foreseen. 

The proposed solution would be particularly useful in countries with a low density of IoT devices as frame retransmissions due to collisions reduce the efficiency of the overall system. Countries that do not enforce the duty cycle regulation (in rural areas, for example) would allow for a more frequent transmission, thus improving the data rate. 
Future work will explore the implementation of machine learning algorithms at the end node, leveraging the embedded computer's capabilities, so that local feature  extraction   will reduce the communication burden and allow for more frequent transmissions.
Still more complex tasks can be assigned to the Application Server, thus implementing the edge computing paradigm.

\section*{Acknowledgment}

This work was partially supported by the ``Conselleria de Educaci\'on, Investigaci\'on, Cultura y Deporte, Direcci\'o General de Ci\'encia i Investigaci\'o, Proyectos AICO/2020", Spain, under Grant AICO/2020/302

\bibliographystyle{IEEEtran} 
\bibliography{bib,bib2}

\end{document}